\documentclass[prl,final,twocolumn,showpacs]{revtex4}

\usepackage{graphicx}
\usepackage{dcolumn}

\begin{document}
\title{Non-equilibrium dynamics of Andreev states in the Kondo regime}
\author{A. Levy Yeyati, A. Mart\'{\i}n-Rodero and E. Vecino} 
\affiliation{Departamento de F\'\i sica de la Materia Condensada C-V. 
Facultad de Ciencias. \\
Universidad Aut\'onoma de Madrid. E-28049 Madrid. Spain.} 
\date{\today}

\begin{abstract}
The transport properties of a quantum dot coupled to superconducting
leads are analyzed. It is shown that the 
quasiparticle current in the Kondo regime
is determined by the non-equilibrium dynamics of
subgap states (Andreev states) under an applied voltage. 
The current at low bias is suppressed exponentially for decreasing Kondo 
temperature in agreement with recent experiments. We also predict novel
interference effects due to multiple Landau-Zener transitions between
Andreev states.
\end{abstract}


\maketitle
   
The interplay between Kondo effect and superconductivity
can lead to a rich phenomenology like the exotic behavior of 
heavy-fermion compounds \cite{cox}.
Artificial devices like quantum dots (QD's) now open the
possibility of exploring this interplay in well characterized transport
experiments \cite{goldhaber}. 
Very recently, Buitelaar et al. \cite{buitelaar} have achieved a 
physical realization of a S-QD-S system using
carbon nanotubes coupled to Al/Au leads.
Their results, showing an enhancement of the
conductance in the Kondo regime with respect to the normal case, have
not hitherto received a theoretical explanation. 

Most theoretical efforts on S-QD-S systems
have been so far devoted to analyze the Josephson
transport in equilibrium conditions 
\cite{varias,vecino}. This system exhibits a  
transition from a Kondo-like to a localized moment regime 
as a function of $T_K/\Delta$,
where $T_K$ is the Kondo temperature in the normal state and $\Delta$ 
is the superconducting energy gap. Associated with this transition the
current-phase relation undergoes a reversal of sign usually known as
$\pi$-junction behavior \cite{varias,vecino}. 

A remarkable feature of this system, which is common to all
coherent nanostructures coupled to superconducting electrodes, is the
appearance of subgap states (usually known as Andreev states (AS's)) 
which give the main contribution to the Josephson current 
\cite{superlattices}. These current
carrying bound states arise from an infinite series of Andreev
reflections at the superconducting electrodes.   
In Ref. \cite{vecino} it was shown that electron correlations in 
a S-QD-S in the
Kondo regime give rise to a renormalization of the AS's which is
consistent with the renormalization of the Fermi liquid parameters in
the normal state. 

In order to obtain theoretical predictions that could be compared with
actual transport experiments one should address the 
situation in which a finite bias voltage is applied. 
This is a challenging theoretical problem in which the current is due to 
multiple Andreev reflection 
(MAR) processes involving strongly correlated electrons. Although some
numerical calculations have already been presented 
\cite{avishai}, a detailed understanding of 
the transport properties in the voltage biased situation is still
lacking.
The problem is particularly involved in the low bias limit in which
the current is due to Andreev processes whose order 
$n \sim int(2\Delta/ev)$ becomes
arbitrarily large as the voltage $v$ tends to zero. 
The aim of this letter is to show that
these properties can be understood in terms of the dynamics of the
Andreev states under an applied bias voltage.

\begin{figure}
\vspace{-0.25cm}
\hspace{-0.5cm} \includegraphics[width=9cm]{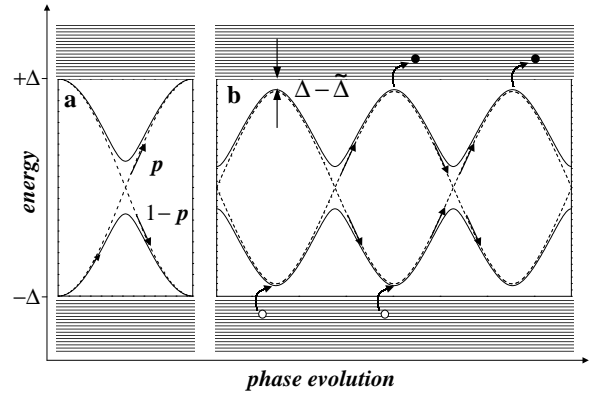}
\vspace{-0.75cm}
\caption{Schematic representation of the Andreev states dynamics in two
different situations: a) quantum point contact and b) S-QD-S in the
Kondo regime. The dashed lines correspond to the states for perfect
normal transmission. The lower and upper striped regions represent the states
in the continuous spectrum. The straight arrows indicate 
Landau-Zener transitions between
AS's, while the curved arrows represent inelastic transitions between
the AS's and the continuous spectrum (see text).}
\end{figure}

In order to give a qualitative explanation of this dynamics
let us first analyze the behavior of
the AS's in different situations, as illustrated in Fig. 1. 
In the absence of a QD, a single channel S-S contact 
with normal transmission $\tau$ exhibits
AS's located at $\pm \Delta \sqrt{1 - \tau \sin^2{\phi/2}}$ \cite{asqpc},
where $\phi$ is the phase drop across the contact. In equilibrium and 
at low temperatures only the lower state is occupied.
Under a small bias voltage this state evolves adiabatically
except for Landau-Zener (LZ) transitions
to the upper state with probability $p = \exp(-\pi \Delta r/ev)$, where
$r = 1 - \tau$ \cite{averin}. This
mechanism produces a net transfer of quasiparticles from the occupied 
to the empty states above the superconducting gap
accounting for the low bias dc current in the system, as represented
by the arrows in Fig. 1 a. As shown below,
in a S-QD-S in the Kondo regime the situation is
qualitatively different due to the fact that the AS's {\it detach} from the
continuous part of the spectrum (see Fig. 1 b).
As a consequence the conversion of bound state electrons into 
free quasiparticles now requires an inelastic transition across the gap
between the AS and the edge of the continuous spectrum.
This feature is the main ingredient which determines the behavior of the
current at low voltages. In particular we will show that there is an
exponential suppression of the current 
as a function of the parameter $T_K/\Delta$. The detachment of the AS's
from the continuous spectrum also gives rise to novel interference
effects due to multiple LZ transitions between the bound
states which could be observed directly as an oscillatory behavior of the
dc current. 

To describe a S-QD-S we start from the usual single level Anderson
model, thus assuming a sufficiently large level separation. 
The model Hamiltonian is $H=$ $H_L$ + $H_R$ + $H_T$
+ $H_D$, where $H_{L,R}$ describe the left and right leads as BCS
superconductors, $H_T = \sum_{k, \sigma, \mu= L,R} 
t_{\mu} \hat{c}^{\dagger}_{k,\sigma} d_{\sigma} + \mbox{h.c}$, 
is the term coupling the dot to the leads,
$\hat{d}^{\dagger}_{\sigma}$ and  $\hat{c}^{\dagger}_{k,\sigma}$
being creation operators for electrons
in the dot and in the leads respectively. 
$H_D = \sum_{\sigma} \epsilon_0 \hat{n}_{\sigma} + U \hat{n}_{\uparrow}
\hat{n}_{\downarrow}$ is the uncoupled dot Hamiltonian
characterized by the dot level position $\epsilon_0$ and the Coulomb
interaction $U$. It is assumed that the coupling of the dot to the leads
in the normal state can be described by energy independent tunneling
rates $\Gamma_L = \Gamma_R = \Gamma$. As discussed in Ref. 
\cite{tocho}
the bias voltage can be introduced as a time-dependent phase factor in
the hopping term.

A crucial point in the analysis of this model is how to deal with the
electron correlation effects. In Ref. \cite{vecino}
a perturbative approach was used to study the zero-voltage case. It was
found that the AS's in the Kondo regime $(T_K > \Delta)$ have essentially
the same phase dependence as in the non-interacting case but with a 
reduced amplitude. This is consistent with a Fermi liquid 
description of the normal state with renormalized 
parameters $\epsilon^*_0$ and $\Gamma^*$, where $\Gamma^* \simeq T_K/2$ fixes
the width of the Kondo resonance. Although different approximations
differ in the way to relate these quantities to the bare 
parameters, the description in terms of the renormalized ones 
can be considered as universal \cite{comment:slave-boson}.
In the opposite limit ($T_K < \Delta$), the system is well described
within the Hartree-Fock approximation \cite{vecino,avishai}. 

In the present work we shall be concerned with the Kondo regime.
At zero bias the AS's are determined by the poles of the dot retarded
Green function, which in the Nambu space and close to the Fermi
energy adopts the form

\begin{eqnarray}
G^r_{D}(\omega) \simeq \frac{\Gamma^*}{\Gamma}
\left( \begin{array}{cc} \omega -
\epsilon^*_0 - 2 \Gamma^* g^r & 2 \Gamma^* f^r \cos{\phi/2} \\
2 \Gamma^* f^r \cos{\phi/2} & \omega + \epsilon^*_0 - 2 \Gamma^* g^r
\end{array}
\right)^{-1} ,
\label{dot-retGF}
\end{eqnarray}
where $g^r = -(\omega+i0^+)/\sqrt{\Delta^2 - (\omega+i0^+)^2}$ and 
$f^r = \Delta/\sqrt{\Delta^2 - (\omega+i0^+)^2}$ correspond to the 
dimensionless Green functions of the uncoupled electrodes. 
In the electron-hole symmetric case ($\epsilon^*_0 = 0$) the AS's are then
determined by the equation

\begin{eqnarray}
\omega_s \pm \Delta \cos{\phi/2} + 
\frac{\omega_s \sqrt{\Delta^2 - \omega_s^2}}{2 \Gamma^*} = 0 .
\label{determinant}
\end{eqnarray}

The solutions of Eq. (\ref{determinant})
for $\Gamma^* \gg \Delta$ and $|\phi| \ll 1$
are well approximated by 
$\omega_s = \pm \widetilde{\Delta} \cos{\phi/2}$ with 
$\widetilde{\Delta} = \Delta[1 - 2(\Delta/2\Gamma^*)^2]$. 
Thus, as commented above, the AS's for the S-QD-S are detached from the
continuous spectrum for finite values of $T_K/\Delta$, with a gap
given by $\Delta - \widetilde{\Delta}$ (see Fig. 1 b).
For decreasing $T_K/\Delta$ the AS's begin to deviate from this simple
phase-dependence \cite{vecino}.

In a non-symmetric situation ($\epsilon^*_0 \neq 0$) the 
two ballistic states are coupled and 
there appear an upper and a lower bound state with an internal gap
of the order of $2 \Delta \sqrt{1-\tau}$, where 
$\tau = 1/\left[1+(\epsilon^*_0/2\Gamma^*)^2\right]$ 
is the normal transmission at the Fermi
energy. This is similar to the behavior of the AS's in a point contact
with finite reflection probability, except for the detachment
from the continuous spectrum (see Fig. 1 b).

In order to analyze the voltage biased situation we shall assume that
the description in terms of the renormalized parameters $\epsilon^*_0$ and
$\Gamma^*$ is still valid in the voltage range ($ev \ll \Delta$)
in which we are interested. Notice that in the Kondo regime this
corresponds to $ev \ll T_K$ and thus one can safely assume that the
coherence associated with the Kondo effect will not be destroyed 
\cite{kaminski}.   

For calculating the current we use a combination of Keldysh and Nambu
formalisms following Refs. \cite{tocho}. Explicit extensions of this
approach to the case of interacting quantum dots are discussed in Refs.
\cite{vecino,us97}.
This procedure allows to evaluate the
current in terms of Green functions  
as $I(t,v) = \sum_n I_n(v) \exp{[in\phi(t)]}$, where $\phi(t) =
2evt/\hbar + \phi_0$. In the present case we shall be interested    
in the dc component $I_0(v)$.

\begin{figure}[ht]
\vspace{-0.5cm}
\includegraphics[width = 8 cm]{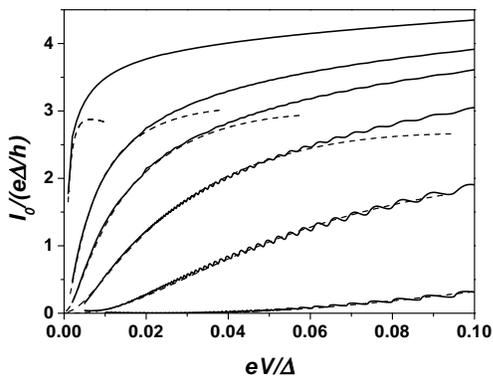}
\caption{Low bias $IV$ characteristics of a S-QD-S in the Kondo regime
for perfect normal transmission and different values of $\Gamma^*/\Delta =
1$, 2, 3, 4, 5 and 10 from bottom to top. The full lines correspond
to the numerical results, while the dashed lines correspond to the
approximation given by Eq. (\ref{asymptotic}).}
\end{figure}

We shall analyze first the electron-hole
symmetric case ($\epsilon^*_0 = 0$). The behavior of the 
dc current at low bias is shown in Fig. 2. 
When $\Gamma^* \gg \Delta$ 
the $IV$ characteristics of a ballistic quantum point contact is
gradually recovered, exhibiting a
divergent low bias conductance with a saturation of the current 
at $v=0$ towards the value $4e\Delta/h$ \cite{averin,tocho}.
As $\Gamma^*/\Delta$ is reduced the low bias conductance is 
gradually suppressed.
When $\Gamma^* \sim \Delta$ oscillations with a period $\Delta/n$, where
$n$ is an integer, starts
to be observable in the $IV$ characteristic. 
These oscillations arise from the opening of new Andreev channels.
As discussed in Ref. \cite{us97}, MAR processes of odd (even) order are
enhanced (suppressed) due to the presence of the dot level at the center
of the gap.     

These results can be understood in more detail
by analyzing the dynamics of the AS's under
a finite bias voltage. In order to establish a dc
current the quasiparticles should be excited from the AS's to the
continuum states. 
A simple diagonalization of Eq.
(\ref{dot-retGF}) for $\epsilon^*_0 = 0$ allows to identify 
$\Sigma(\omega) = \omega \sqrt{\Delta^2-\omega^2}/2\Gamma^*$ 
as a complex self-energy for the ballistic states at $\pm \Delta
\cos{\phi/2}$. 
The real part of $\Sigma(\omega)$ within the gap is
responsible for the renormalization of the bound states for 
finite $\Gamma^*$
(see Eq. (\ref{determinant})). When a small bias voltage is applied 
the AS's tend to follow adiabatically the 
evolution of the phase. The transition rate into the continuum 
is determined by the imaginary part of the self-energy and
can be calculated using standard time-dependent
perturbation theory. The probability for this transition
is then given by 

\begin{eqnarray}
P(t)  & = & \frac{1}{\hbar^2}  \int_\Delta^{\infty} 
d\omega \mbox{Im} \Sigma(\omega) \, \times \nonumber\\
& & \left|\int_0^{t} dt_1 e^{
i\int_0^{t_1} dt_2 \left(\omega - v/2 - \omega_s(t_2) 
\right)/\hbar}\right|^2 .
\end{eqnarray}

Using the approximate form of the AS, $\omega_s \simeq
\widetilde{\Delta} \cos{\phi/2}$ (valid for large $\Gamma^*/\Delta$) and
in the stationary limit ($t \rightarrow \infty$)
one obtains the rate for this transition, $\gamma_T = \sum_{n\ge0} \gamma_n$, with

\begin{eqnarray}
\gamma_n = \frac{2 \pi}{\hbar}
J_n^2(\widetilde{\Delta}/v)
\mbox{Im} \Sigma(nv+v/2) 
\theta(nv+v/2-\Delta)  ,
\label{rate}
\end{eqnarray}
where $J_n$ are the integer order Bessel functions and $\theta$ is the
step function. In this decomposition the rates $\gamma_n$ 
correspond to processes in which $(2n+1)e$ are transferred,  
which allows to write the dc current as
$I_0 = e \sum_n (2n+1) \gamma_n$.
In the limit $v \rightarrow 0$, $nv+v/2$ in Eq. (\ref{rate}) can be
taken as a continuous variable $x$. From the asymptotic behavior of
$J_n$ for large $n$ one can obtain a simplified expression for the 
current at low bias 

\begin{eqnarray}
I_0 = \frac{2e}{\Gamma^* v \hbar} 
\int_{\Delta}^{\infty} dx x^2 
\sqrt{\frac{x^2 - \Delta^2}{x^2 -
\widetilde{\Delta}^2}} e^{2x(\tanh{\alpha} - \alpha)/v} ,
\label{asymptotic}
\end{eqnarray}
where $\cosh{\alpha} = x/\widetilde{\Delta}$. 

The predictions of Eq. (\ref{asymptotic}) are compared with the 
full numerical results
in Fig. 2. It should be stressed that the asymptotic expression is 
only valid in the voltage range $ev \leq (\Delta - \widetilde{\Delta})$.
Within this range the current is suppressed according to Eq.
(\ref{asymptotic}), decreasing exponentially with the parameter $(\Delta
- \widetilde{\Delta})/ev$. The agreement with the full numerical results
support our interpretation in terms of the AS's dynamics. 

In the absence of electron-hole symmetry ($\epsilon^*_0 \neq 0$) the
dynamics of the AS's become more complex due to interference effects.
To establish a dc current now requires 
LZ transitions between the lower and the upper state 
in addition to the mechanism previously
discussed. In contrast to the case of a quantum point contact (in which
the AS's merge with the continuous spectrum at $\phi = 2n \pi$), 
in the S-QD-S case there is a finite probability of preserving the
system coherence after a complete cycle. As a consequence interference
effects between subsequent LZ transitions may appear. 
The two main paths for this interference correspond to the evolution
following the lower and upper AS, as schematically shown in
Fig. 1 b. The phase difference between these two paths is thus simply
given by $\int_\pi^{3\pi} d\phi \, \omega_s(\phi)/ev$. One would 
therefore
expect oscillations of the form $\cos^2(2\widetilde{\Delta}/ev)$ to be
observable in the current in addition to
the ones associated with the onset of new MAR processes.  
 
\begin{figure}[h]
\vspace{-0.25cm}
\includegraphics[height=6cm]{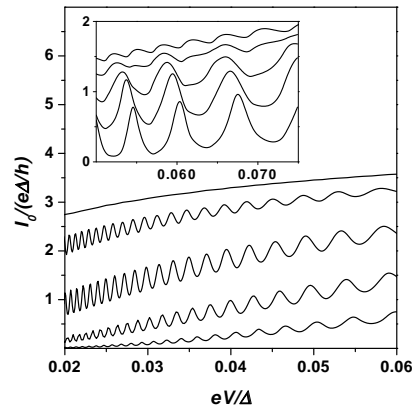}
\caption{Evolution of the oscillations pattern in the $IV$
characteristics as a function of $\epsilon^*_0$ for $\Gamma^*/\Delta = 5$ and 
$\epsilon^*_0/\Delta = 0$, 0.5, 1.0, 1.5 and 2 from top to bottom. Inset:
$\Gamma^*/\Delta = 2$ and $\epsilon^*_0/\Delta = 0$, 0.2, 0.4, 0.6 and 0.8.} 
\end{figure}

Fig. 3 shows the numerical results for the low bias
dc current in a non-symmetric
situation. 
As can be observed, when increasing $\epsilon^*_0$ there is an overall
suppression of the current together with the onset of an oscillatory  
behavior of increasing amplitude. These curves correspond to a large
$\Gamma^*/\Delta$ value in order to disentangle these oscillations
from the MAR induced ones. The overall suppression is governed by the LZ
probability $p$ \cite{averin}. 
The superimposed oscillations exhibit the expected behavior associated
with the interference between AS's due to multiple LZ transitions. 
A simple argument suggests that the amplitude of these oscillations 
should scale as $p(1-p)$ and therefore be maximum for $p=1/2$ (i.e.
$ev/\Delta \simeq \pi r/\ln 2$),
as confirmed by the numerical results.
When $\Gamma^*/\Delta$ is reduced one can observe a superposition 
of the oscillations arising from the two mechanisms discussed above. The
inset of Fig. 3 illustrates the transition from one type of oscillations
to the other in a reduced voltage range.

\begin{figure}[h]
\includegraphics[width = 7 cm]{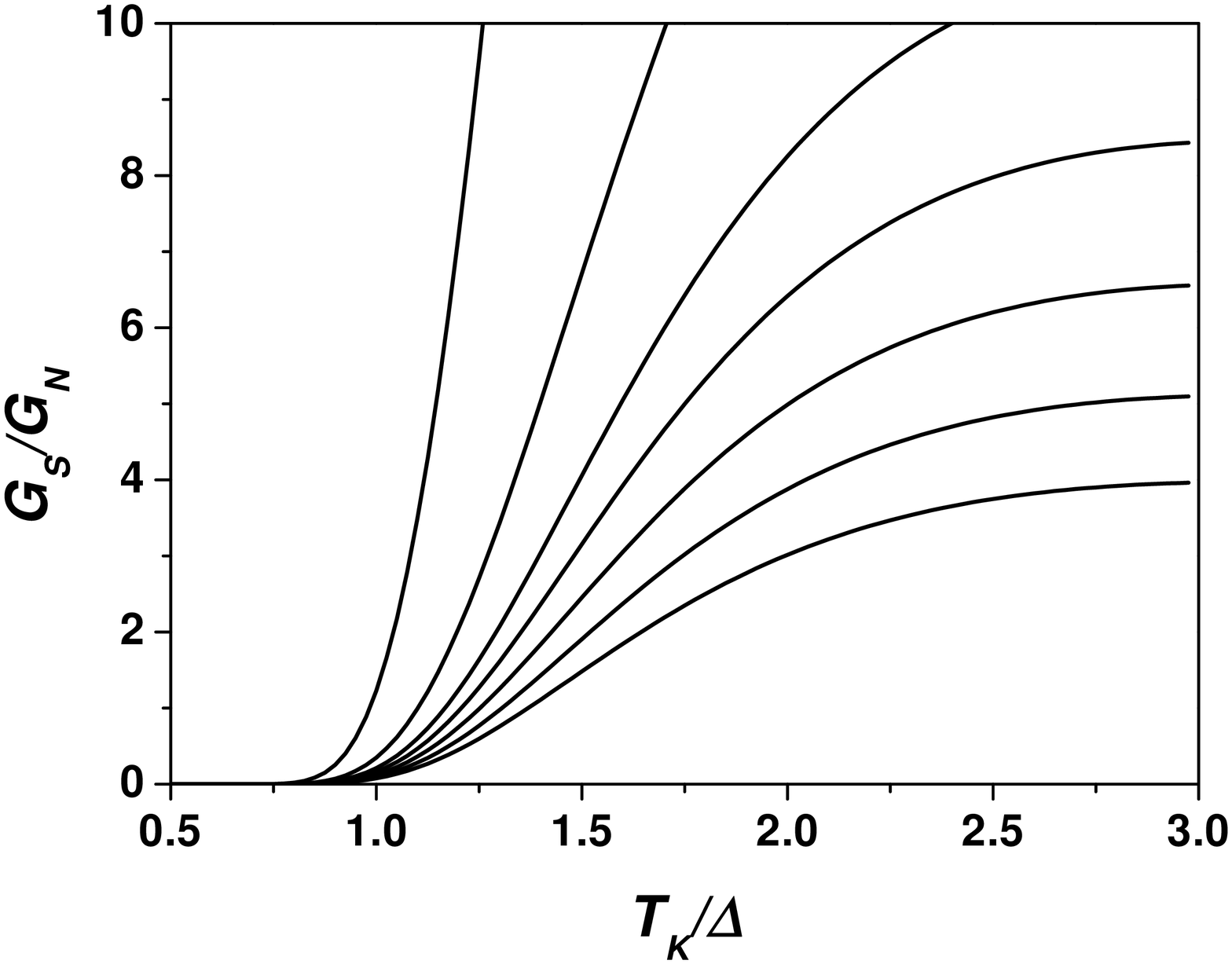}
\caption{Zero-bias conductance defined by introducing a finite voltage
resolution $e\delta v/\Delta = 0.025$. The normal transmission from
bottom to top is 0.978, 0.980, 0.982, 0.984, 0.986, 0.99 and 1.0.}
\end{figure}

The results presented above indicate that the $IV$ curves of a S-QD-S in
the Kondo regime are extremely non-linear in the low bias limit and thus
the system cannot be characterized just by its zero-bias conductance. 
In an actual experiment the observability of the predicted behavior
will be limited by several possible factors like the multilevel
structure of the QD, environmental effects and inelastic relaxation 
processes which could be of importance at finite temperature.
In the present calculations we have assumed sufficiently small
temperatures so that the inelastic relaxation is negligible 
within the voltage range considered. In order to attempt a 
comparison with the results of Ref. \cite{buitelaar} one should 
take into account that the actual voltage resolution is limited to some 
minimum value $\delta v$ in such a
way that the measured zero-bias conductance is $G_S \simeq 
I_0(\delta v)/\delta v$.
In these experiments $\delta v$ can be estimated to be
around $2.5 \mu V$, which corresponds to $e\delta
v/\Delta \simeq 0.025$ \cite{private}.  
The zero-bias conductance defined by this finite voltage resolution
is plotted in Fig. 4 as a function of $T_K/\Delta$.
When $T_K/\Delta \gg 1$ the conductance $G_S$ tends to saturate to 
a value which can be several times larger than the normal 
conductance. The actual value depends on 
$\delta v$ and is very sensitive to slight deviations from 
perfect transmission (deviations as small as $2 \%$ from $\tau=1$
produce variations by a factor $\sim 10$ in $G_S/G_N$). 
On the other hand, the conductance
exhibits an exponential suppression 
for $T_K/\Delta \leq 1$. 
All these predictions are in qualitative agreement with the results 
presented in Fig. 4 of Ref. \cite{buitelaar}.

In conclusion we have shown that the transport properties of a
S-QD-S in the Kondo regime can be understood from the dynamics of
the Andreev states under an applied bias voltage. The quasiparticle
current at low bias is controlled by inelastic transitions between the
AS's and the extended states in the continuous spectrum. 
On the other hand, we predict novel interference
effects due to multiple Landau-Zener transitions between Andreev states
which could be worth to explore in future experiments.

We thank C. Sch\"onenberger and M.R. Buitelaar
for useful comments on their experiments and also J.C. Cuevas, 
R. Aguado and C. Urbina for fruitful discussions. 
One of us (E.V.) would like to thank J.C. Cuevas and G. Sch\"on for their
hospitality during his stay at the Karlsruhe Institut f\"ur Theoretische
Festk\"orperphysik.
This work has been supported by grant BFM2001-0150 (MCyT, Spain).


\begin{thebibliography}{}
\bibitem{comment:slave-boson} An equivalent description in terms of
renormalized parameters is obtained
within the slave-boson mean field approach. See for instance last reference
in \cite{avishai}.
\end{thebibliography}
\end{document}